\documentclass[runningheads]{llncs}
\usepackage[T1]{fontenc}
\usepackage{graphicx}
\usepackage{booktabs}
\usepackage{array}
\usepackage{graphicx}
\usepackage{tikz}
\usepackage{booktabs}
\usepackage{siunitx}
\usepackage{multirow}
\usepackage[title]{appendix}%
\usepackage{xcolor}
\usepackage{caption}

\begin{document}
\title{Should LLM Agents Decide in Social Simulations? Comparing Finite-State and LLM-Based Decision Policies}
\titlerunning{Should LLM Agents Decide?}
%
\author{Alejandro Buitrago L\'opez\inst{1}\orcidID{0009-0002-1606-8766} \and
Javier~Pastor-Galindo\inst{1}\orcidID{0000-0003-4827-6682} \and
Jos\'e~A.~Ruip\'erez-Valiente\inst{1}\orcidID{0000-0002-2304-6365}}
\authorrunning{Buitrago L\'opez et al.}
%
\institute{
Dept. of Information and Communications Engineering, University of Murcia, 30100, Spain\\
\email{\{alejandro.buitragol,javierpg,jruiperez\}@umu.es}}
\maketitle              

\begin{abstract}
Large language models (LLMs) are increasingly used as decision-making components in social simulations. This introduces a methodological risk: the simulation may deviate from the explicit behavioral policy defined by the researcher. In online social network (OSN) simulations, action choices shape system dynamics, interaction patterns, and model interpretability.

This paper evaluates whether LLM action selectors preserve an interpretable reference policy in an OSN simulation. The reference is a finite state machine implemented as a first-order Markov model, with transition probabilities depending on the user type. The evaluation uses a synthetic network with $1{,}000$ agents and $10{,}000$ action decisions. Three open-weight LLMs are tested: LLaMA 3.1, GPT-OSS, and Mistral 24B. Each model is evaluated under three prompting strategies: base, guided, and probabilistic. Alignment is measured using Jensen--Shannon Divergence with Laplace smoothing, and execution time is reported. Results show that LLMs can approximate the reference policy in some configurations, but do not preserve it reliably. Alignment varies across models and prompts, and additional guidance can introduce systematic action biases. Even the best-aligned LLM configurations are several hundred times slower than direct Markov chain sampling. These findings indicate that LLM-based action selection is not a direct replacement for explicit decision policies: it can alter the intended behavior while increasing computational cost.

\end{abstract}
\section{Introduction}

Agent-based simulation is widely used to study complex social systems because it makes agents, interactions, and decision rules explicit, enabling researchers to examine the collective patterns that emerge from them \cite{epstein2006generative}. In online social network (OSN) simulations, this is particularly important because exposure, interaction, diffusion, and network evolution all depend on the actions selected by individual users \cite{10492674}. As a result, the validity of the simulation depends on how those actions are chosen: if the decision policy changes, the behavior of the system may also change.

Traditional OSN simulations often rely on explicit behavioral rules, finite-state machines, or probabilistic transition policies \cite{lopez2024survey}. These mechanisms simplify human behavior, but their assumptions are inspectable, their outputs are reproducible, and their computational cost is low. This is especially important in adversarial or risk-oriented simulations, where small changes in action selection can alter measured reach, engagement, diffusion, or campaign amplification \cite{nasim2025influence}. If the decision policy is opaque, it becomes difficult to determine whether an observed outcome is due to the simulated scenario or to hidden biases in the mechanism controlling the agents.

LLM-based agent simulations are increasingly used to model agents capable of interpreting context, using memory, interacting with others, and generating language \cite{yang2024oasis}. In many cases, however, the LLM is not only used to generate content (e.g., messages or replies), but also to decide the agent’s next action. In an OSN simulation, deciding which actions to perform (e.g., reading, posting, or following) defines both agent behavior and the evolution of the simulation. When this decision is delegated to an LLM through prompting, the behavioral policy becomes dependent on the model, the prompt, and the context, which may introduce non-transparent biases and variability across settings \cite{10.1145/3774905.3795477}. Therefore, the key question is not only whether LLM agents can produce plausible content, but whether their action choices preserve the intended decision policy.

In this paper, we study this problem by comparing two action selection mechanisms within the same OSN simulation. The first is an explicit and interpretable finite-state decision policy, where agents move between social-media actions according to user-type-specific transition probabilities. The second delegates the same decision to an LLM through prompting. This comparison tests whether an LLM-based selector preserves the intended behavioral policy or introduces a different action distribution. The research question is: \emph{to what extent do LLM-based action choices preserve an interpretable finite-state decision policy, and how does this alignment change across models and prompting strategies?}

To conduct this evaluation, we use an agent-based OSN framework \cite{buitrago2025agentbased}, where agents perform common social-media actions: \textit{read}, \textit{like}, \textit{share}, \textit{reply}, \textit{post}, \textit{follow}, and \textit{unfollow}. The paper makes three main contributions:

\begin{itemize}
    \item A controlled comparison is performed between two action-selection approaches under identical agents, network, action space, and simulation context: an explicit FSM/Markov policy based on user-type-specific transition matrices, and LLM-based selectors that choose one valid action through prompting.
    \item The evaluation is conducted in a synthetic OSN with 1,000 agents and 10,000 action-selection steps, using three open-weight LLMs (LLaMA 3.1, GPT-OSS, and Mistral 24B) and three prompting strategies.
    \item Alignment with the FSM is measured using Jensen--Shannon Divergence (JSD) with Laplace smoothing, and the execution cost of each decision policy is reported.
\end{itemize}

\section{Related work}
\label{sota}

Agent-based simulation has traditionally used explicit rules, finite-state machines, probabilistic policies, or Markov-based models to define how agents behave \cite{epstein2006generative,norris1997markov,lopez2024survey}. These approaches simplify human behavior, but their assumptions are inspectable, reproducible, and computationally efficient. This is important in OSN simulations, where user actions shape exposure, diffusion, and network structure \cite{yang2024oasis,liu2025mosaic}. For example, the frequency of sharing affects diffusion \cite{bakshy2011everyone}, while following and unfollowing affect network structure.

LLM-based agents have expanded this setting. LLMs enable agents to interpret context, use memory, generate language, plan behavior, and interact with other agents \cite{park2023generative}. This makes simulations more flexible, but it also changes how agent behavior is controlled. In some systems, the LLM is used mainly to generate text. In others, it also decides what the agent should do next. When action selection is delegated to an LLM, the resulting behavior becomes dependent not only on explicit rules, but also on model-specific responses to prompts and contextual inputs.

Several recent simulators give LLMs a direct role in agent behavior. OASIS uses LLM agents in a social media environment where agents receive platform context and produce actions within a broad OSN action space \cite{yang2024oasis}. Y Social proposes an LLM-powered social media digital twin where agents interact through actions such as reading, posting, commenting, sharing, reacting, and following \cite{rossetti2024ysocialllmpoweredsocial}. MOSAIC models social media behavior with agents that choose actions such as liking, sharing, or ignoring posts \cite{liu2025mosaic}. 

Other works use hybrid forms of control. In rumor-spreading simulations, the LLM generates posts and updates agents' beliefs, while external activation strategies decide which agent acts at each step \cite{hu2025rumor}. In influence-dynamics simulations, LLM agents generate persuasive or corrective messages, while opinion changes are governed by explicit update rules \cite{nasim2025influence}. A similar distinction appears in recent work on AI-enabled influence operations in synthetic social networks, where agent behavior is controlled through deterministic state machines and the LLM is used for organic content generation and bounded analytic tasks \cite{buitrago2026influenceops}. These designs retain partial control over the simulation while delegating specific components to the LLM, separating explicit decision mechanisms from generative behavior.

Empirical studies of LLM-driven social behavior suggest that model outputs should not be assumed to match the intended behavior. Evidence from Chirper.ai, LLM-generated reactions to Spanish online news, and survey-based evaluations shows that outputs can differ from human behavior and depend strongly on prompting and configuration \cite{zhu2026characterizingllmdrivensocialnetwork,buitrago2026realism,10.1145/3774905.3795477}. These findings do not invalidate LLM-based simulations, but they do show that observed outcomes may reflect the implicit decision policy induced by the model rather than the behavior intended by the simulator.

The reviewed works can be grouped into three broad forms of simulation control. First, explicit approaches use rules, finite-state machines, probabilistic policies, or Markov-based models \cite{epstein2006generative,norris1997markov,lopez2024survey}. Second, three reviewed LLM-based systems delegate part of agent behavior directly to the LLM, including action selection, planning, or reaction to the environment: OASIS, Y Social, and MOSAIC \cite{yang2024oasis,rossetti2024ysocialllmpoweredsocial,liu2025mosaic}. Third, three works use hybrid control, where LLM-generated content or analysis is combined with explicit activation rules, workflows, belief updates, or opinion dynamics \cite{hu2025rumor,nasim2025influence,buitrago2026influenceops}.

Despite these advances, most existing work evaluates complete simulation outcomes, so the specific behavior induced by the LLM action-selection policy often remains implicit. As a result, it remains unclear whether LLM-based agents reproduce the intended behavioral assumptions or implicitly implement a different policy. This work studies this problem in a controlled OSN setting. The same simulation is run under two forms of action control. In one case, agents choose actions through an explicit FSM/Markov policy based on transition probabilities. In the other, the next action is chosen by an LLM through prompting. Building on a framework where action selection and language generation are separated \cite{buitrago2025agentbased}, the study compares both approaches under the same agents, network, context, and action space. The goal is to quantify how LLM-based action selection aligns with or deviates from an explicit reference policy under controlled conditions.

\section{Experimental setup} \label{sec:met}

\begin{figure}[ht!]
\centering
\includegraphics[width=\linewidth]{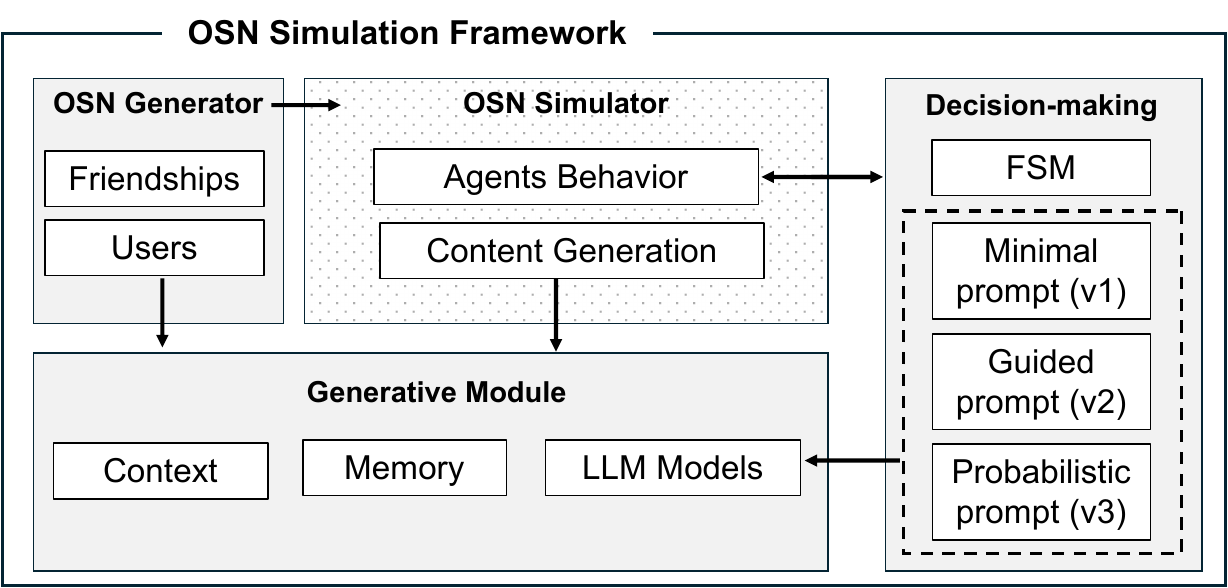}
\caption{Overview of the experimental setup. The OSN generator creates the synthetic users and friendships. The OSN simulator activates agents sequentially. At each step, the activated agent selects one action using either the FSM baseline or one of the three LLM-based prompting strategies.} \label{fig:methodology}
\end{figure}

Figure~\ref{fig:methodology} summarizes the experimental setup. All conditions are evaluated on the same synthetic OSN, with the same agents, network structure, action space, and context-construction procedure. The only component that changes is the action-selection policy: an explicit FSM/Markov baseline or one of three LLM-based policies defined by different prompting strategies. The resulting action distributions are compared with the FSM reference using JSD. Laplace smoothing with $\alpha=10^{-6}$ is applied before normalization to avoid zero-count artifacts. JSD is computed with base-2 logarithms, so values lie in $[0,1]$, where lower values indicate closer alignment with the FSM reference policy.

\subsection{OSN simulation environment and action selection}
\label{sec:synthetic_osn}

The evaluation is conducted using an agent-based OSN simulation framework \cite{buitrago2025agentbased}, representing a microblogging platform with synthetic users, directed social links, and a fixed action space. The simulation follows three stages: agent generation, network construction, and sequential interaction.

Agents are generated with structured profiles including demographic and behavioral attributes (e.g., age, personality, occupation, interests, and user type), which define their semantic characteristics. The social graph is constructed using a homophily-based generation process, following the same framework described in \cite{11441429}, and remains fixed across all conditions. The simulation proceeds through sequential agent activations for a configurable number of steps. At each step, an agent is activated, its local context is derived from the current network state, and one valid action is selected from the common action space.

\[
\mathcal{A} = \{\textit{read}, \textit{like}, \textit{share}, \textit{reply}, \textit{post}, \textit{follow}, \textit{unfollow}\}.
\]

The selected action is executed and updates the simulation state. For example, \textit{read} exposes the agent to content; \textit{like}, \textit{share}, and \textit{reply} require available content; \textit{post} creates new content; and \textit{follow} or \textit{unfollow} modify the social graph when valid candidates exist. All experimental conditions share identical simulation dynamics; only the decision policy differs. The following subsection defines the four action-selection policies compared in this study.

\subsection{Action-selection policies}
\label{sec:decision_policies}

Four action-selection policies are compared. The first one is the original FSM-based decision policy of the simulator. The other three replace the FSM decision with an LLM-based action selector under different prompting strategies. All policies receive the same agent profile, current state, valid actions, and local simulation context.

\subsubsection{FSM baseline}
\label{sec:fsm_policy}

The FSM is the original action-selection mechanism of the simulator \cite{buitrago2025agentbased} and serves as the interpretable reference policy. It represents agent behavior as transitions between discrete OSN actions. Each state corresponds to an action, and the next action is sampled according to a transition matrix. The transition probabilities depend on the current state and the agent's user type. Formally, given a current action state $a_i$ and a user type $u$, the FSM defines a probability distribution over the next action $a_j$:

\[
p^{(u)}_{ij} = P(A_{t+1}=a_j \mid A_t=a_i, U=u),
\qquad
\sum_j p^{(u)}_{ij}=1.
\]

\subsubsection{Base prompt (Prompt v1)}
\label{sec:prompt_v1}

Base prompt (v1) provides the LLM with the agent metadata, recent state, platform constraints, available actions, and short definitions of each action. It includes only minimal behavioral framing to avoid two potential artifacts: framing the task as an optimization problem and artificially balancing actions. This prompt captures the model's default action preferences under a constrained OSN decision context, without user-type-specific calibration or numerical transition probabilities.

\subsubsection{Guided prompt (Prompt v2)}
\label{sec:prompt_v2}

Guided prompt (v2) extends the base prompt with explicit behavioral guidance derived from empirical observations of OSN use. This guidance acts as an explicit intervention on the decision process rather than neutral background information. It summarizes five decision rules: users normally consume available feed content before reacting to it \cite{benevenuto2012characterizing}; content creation is unequally distributed across users \cite{antelmi2019characterizing}; actions differ in effort, from low-cost signals such as \textit{like} to higher-initiative actions such as \textit{reply} and \textit{post} \cite{bakshy2011everyone}; user roles shape engagement patterns \cite{uddin2014understandingtypesuserstwitter}; and relational actions such as \textit{follow} and \textit{unfollow} should be treated as ordinary candidates when available.

The prompt includes calibration rules discouraging the model from over-selecting actions that are easier to justify linguistically, such as \textit{post} or \textit{reply}, and from ignoring lower-effort or relational actions such as \textit{like}, \textit{share}, \textit{follow}, or \textit{unfollow}. This specificity is important for interpreting the results: if guided prompt (v2) induces a strong shift toward a particular action, this reflects how a model applies the provided behavioral rules, not only its baseline preference.

\subsubsection{Probabilistic prompt (Prompt v3)}
\label{sec:prompt_v3}

Probabilistic prompt (v3) gives the LLM the numerical information used by the FSM to choose the next action. For each decision, the model receives the fixed action order, the current state, the transition probabilities for the corresponding user type, the contextual mask, the final normalized probabilities, the allowed actions, and short validity notes. The mask reflects the current situation of the agent, for example whether the agent has already read a post, whether \textit{like}, \textit{share}, or \textit{reply} are possible, whether there are concrete \textit{follow} or \textit{unfollow} candidates, and whether repeated actions such as \textit{post} or \textit{reply} should be downweighted.

The model must choose one action from the allowed set. It is instructed to use the final probabilities as strong behavioral priors, but not to always select the highest-probability action, ignore the probabilities, or artificially balance the actions. The prompt also includes short action definitions and user-type calibration rules, so the model can interpret the numerical policy within the current OSN context. Thus, Probabilistic prompt (v3) provides numerical priors derived from the FSM while leaving final action selection to the model.

Table~\ref{tab:prompt_excerpts} shows representative content and word lengths for the three prompt templates. For probabilistic prompt (v3), the probability values are illustrative; actual values vary across decision instances. The full prompts are provided in the supplementary material.

\begin{table}[t]
\centering
\caption{Representative content and length of the three prompting strategies.}
\label{tab:prompt_excerpts}
\scriptsize
\setlength{\tabcolsep}{3.5pt}
\renewcommand{\arraystretch}{1.12}
\begin{tabular}{p{0.08\linewidth} p{0.18\linewidth} p{0.66\linewidth}}
\toprule
\textbf{Words} & \textbf{Prompt} & \textbf{Representative content} \\
\midrule
213 &
\textbf{Base prompt (v1)} &
\emph{Decide as a realistic human social media user, not as an optimizer. [...] Do not try to maximize action diversity. [...] Prefer the action that feels most natural and plausible right now.} \\
\midrule
670 &
\textbf{Guided prompt (v2)} &
\emph{When readable feed content exists, users normally consume content before reacting. [...] Follow and unfollow are ordinary relationship-management actions. [...] Do not choose post merely because the user is Advanced, Debater, or Socializer.} \\
\midrule
538 &
\textbf{Probabilistic prompt (v3)} &
\emph{Final normalized probabilities after masking, e.g., read=0.55, like=0.20, share=0.08, reply=0.07, post=0.05, follow=0.05, unfollow=0.00. [...] Use the final normalized probabilities as strong behavioral priors. Do not simply choose the highest-probability action every time.} \\
\bottomrule
\end{tabular}
\end{table}

\section{Experimental results}
\label{sec:results}

This section evaluates how action distributions change when the FSM-based decision policy is replaced by LLM-based action selection under controlled conditions. All models are tested within the same experimental setup: a synthetic OSN with 1,000 agents divided into four behavioral profiles: 547 Passive users (54.7\%), 221 Socializers (22.1\%), 135 Debaters (13.5\%), and 97 Advanced users (9.7\%). Each configuration consists of 10,000 action-selection steps with identical action space, decision context-construction procedure, prompts, and decoding settings.

We compare three open-weight LLMs: \texttt{Llama-3.1-8B}, \texttt{gpt-oss-20b-BF16}, and \texttt{Mistral-Small-3.2-24B}. These models were selected to cover different open-weight model families while allowing deployment under the same local inference setup. This design tests whether prompt-induced effects are consistent across models rather than specific to a single implementation. All experiments were conducted on a dedicated Ubuntu server with $10$ CPU cores (Intel Xeon, $2.3$ GHz) and $100$ GB of RAM. The models were served through a local vLLM endpoint. For each decision instance, the model received the corresponding decision context and was required to return exactly one valid action from the predefined action space. All requests use the same decoding configuration: \texttt{temperature}=0.7, \texttt{top\_p}=0.9, and \texttt{top\_k}=40.

\subsection{LLaMA 3.1}
\label{sec-llama}

\begin{figure}[ht!]
\centering
\includegraphics[width=\linewidth]{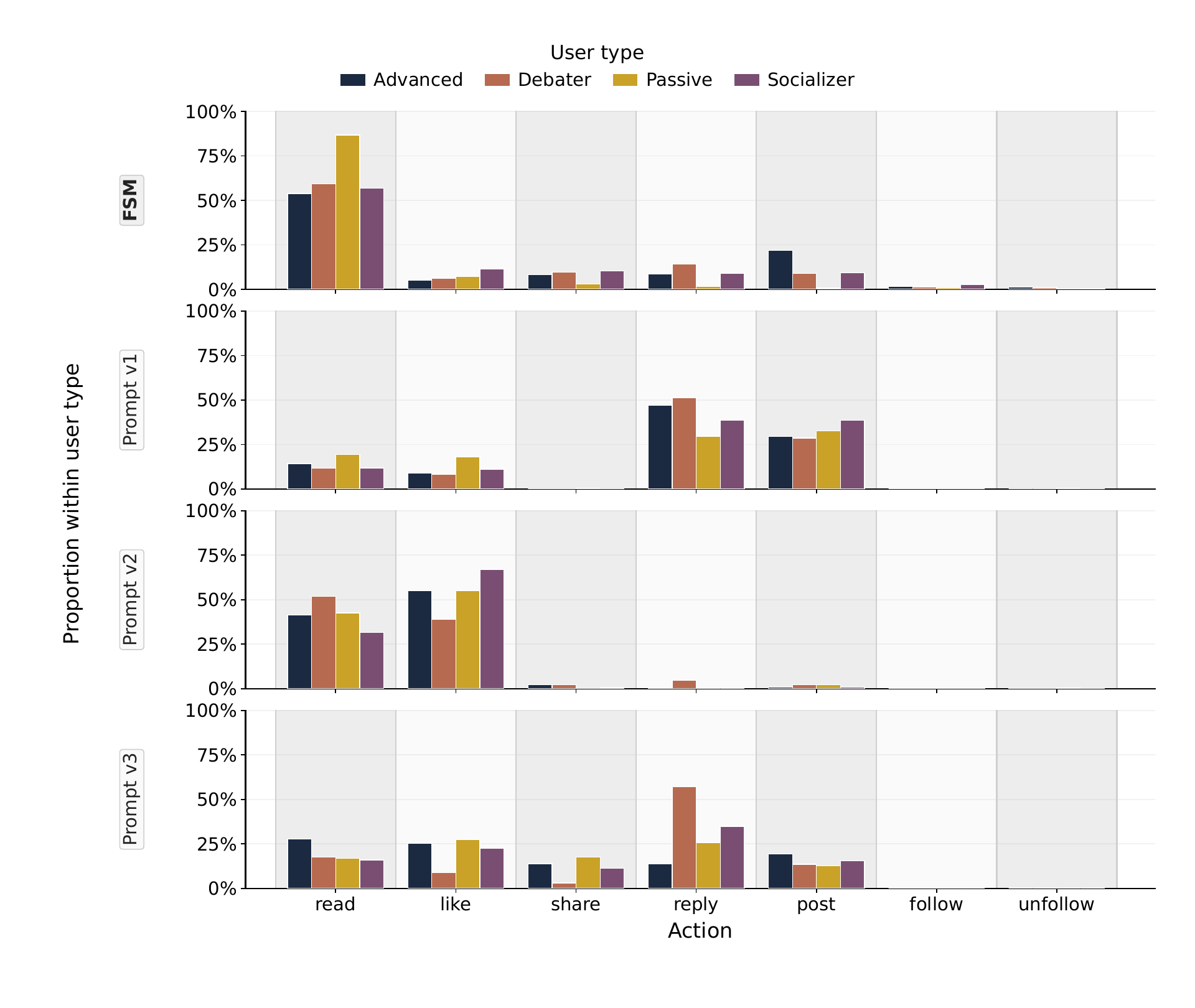}
\caption{Action distributions produced by LLaMA 3.1 under the three prompting strategies, compared with the FSM baseline. Bars show the proportion of each action within each user type.}
\label{fig:barllama}
\end{figure}

Figure~\ref{fig:barllama} reports the action proportions produced by LLaMA 3.1 for each user type and prompt, with the FSM shown as the reference distribution. The rows correspond to the FSM baseline and the three prompting strategies, while the columns show the seven possible actions. In the FSM row, \textit{read} is the dominant action across all user types. The remaining actions appear with lower proportions, with \textit{post}, \textit{reply}, \textit{like}, and \textit{share} appearing at lower but non-zero levels across profiles.

The three LLaMA 3.1 prompt rows show substantial differences in the distribution of actions. Under base prompt (v1), the bars for \textit{read} are much smaller than in the FSM row, while \textit{reply} and \textit{post} become the most prominent actions. Guided prompt (v2) reduces the over-selection of \textit{reply} and \textit{post}, but it remains different from the FSM profile. Instead, it increases the proportion of \textit{like}, especially among Passive users and Socializers. Probabilistic prompt (v3) produces a more diverse distribution for some user types and partially restores \textit{share}, but it still differs from the FSM baseline, particularly for Passive users, where the largest deviations are observed.

\begin{figure}[ht!]
\centering
\includegraphics[width=\linewidth]{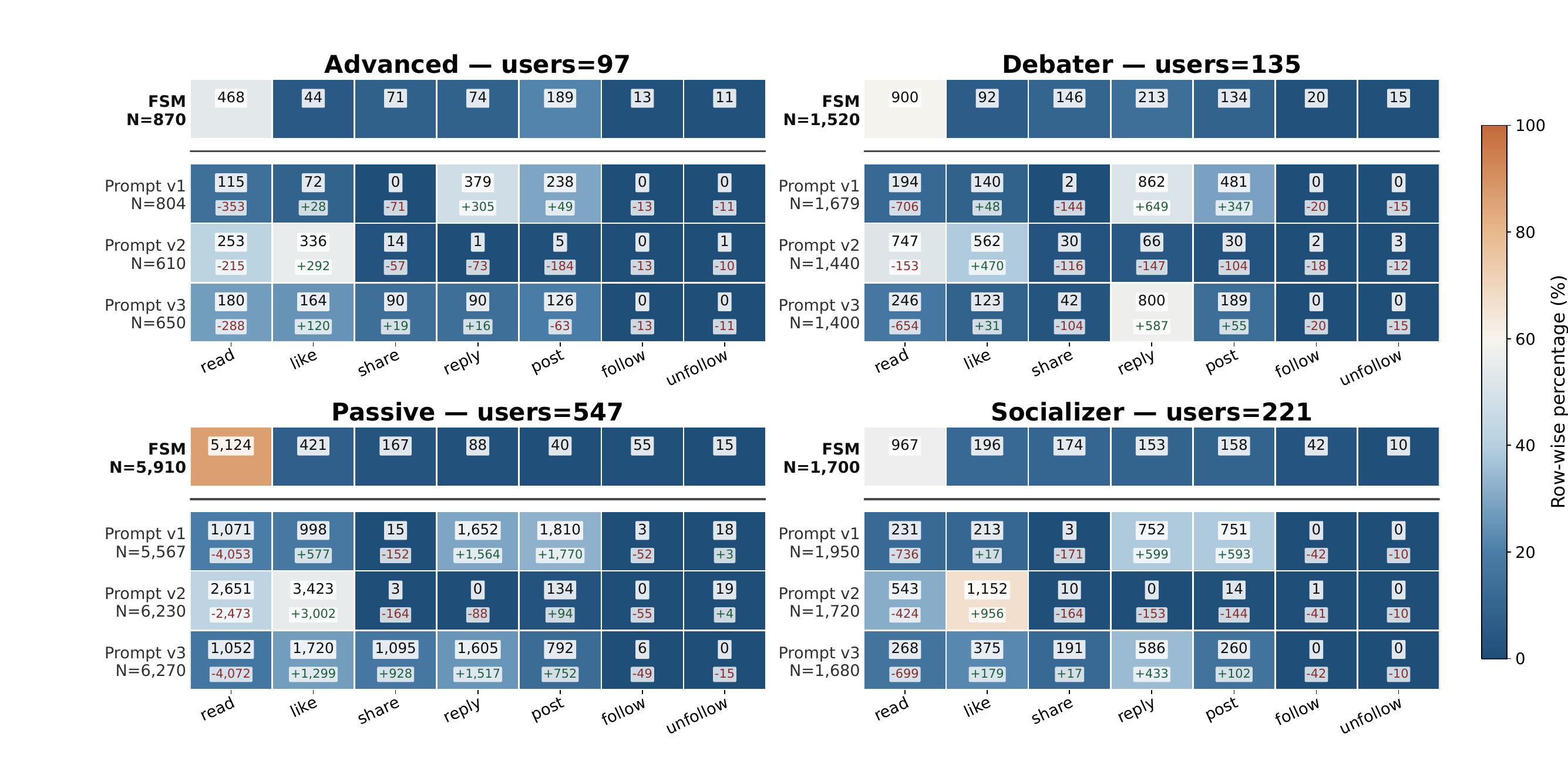}
\caption{Action counts produced by LLaMA 3.1 by user type and prompt. Cell values report raw counts, while secondary values indicate the difference with respect to the FSM baseline for the same user type.}
\label{fig:heatllama}
\end{figure}

Figure~\ref{fig:heatllama} reports the action counts produced by LLaMA 3.1 for each user type. The figure is organized into four panels, one for each user profile. Within each panel, rows correspond to the FSM baseline and the three prompts, while columns correspond to the seven actions. Each cell reports the raw action count; for the LLM rows, the value below the count indicates the difference with respect to the FSM for the same user type and action. The color scale represents the row-wise percentage, so lighter cells identify the actions that dominate each policy.

The heatmap shows differences between LLaMA 3.1 and the FSM that extend across multiple actions rather than being confined to individual cases. Each prompt produces a distinct distribution pattern. Base prompt (v1) moves decisions away from \textit{read} and toward more active actions, especially \textit{reply} and \textit{post}. Guided prompt (v2) reduces this shift, but concentrates a large proportion on \textit{like}. Probabilistic prompt (v3) distributes decisions across more action types and partially restores some lower-frequency actions, although it still differs from the FSM pattern, particularly for Passive users.

JSD results confirm this pattern. At the global level, guided prompt (v2) achieves the closest alignment with the FSM ($JSD=0.223$), followed by probabilistic prompt (v3) ($JSD=0.278$) and base prompt (v1) ($JSD=0.359$). The weighted user-type averages show the same ordering, with guided prompt (v2) obtaining the lowest divergence ($JSD=0.243$) and base prompt (v1) the highest ($JSD=0.390$). However, this global improvement is not uniform across user types. Probabilistic prompt (v3) achieves the lowest divergence for Advanced users ($JSD=0.108$) and Socializers ($JSD=0.181$), whereas guided prompt (v2) performs best for Debaters ($JSD=0.158$) and Passive users ($JSD=0.234$). Probabilistic prompt (v3) shows higher divergence for the Passive profile, where divergence remains high ($JSD=0.420$).

\subsection{GPT-OSS}
\label{sec-gpt}

\begin{figure}[ht!]
\centering
\includegraphics[width=\linewidth]{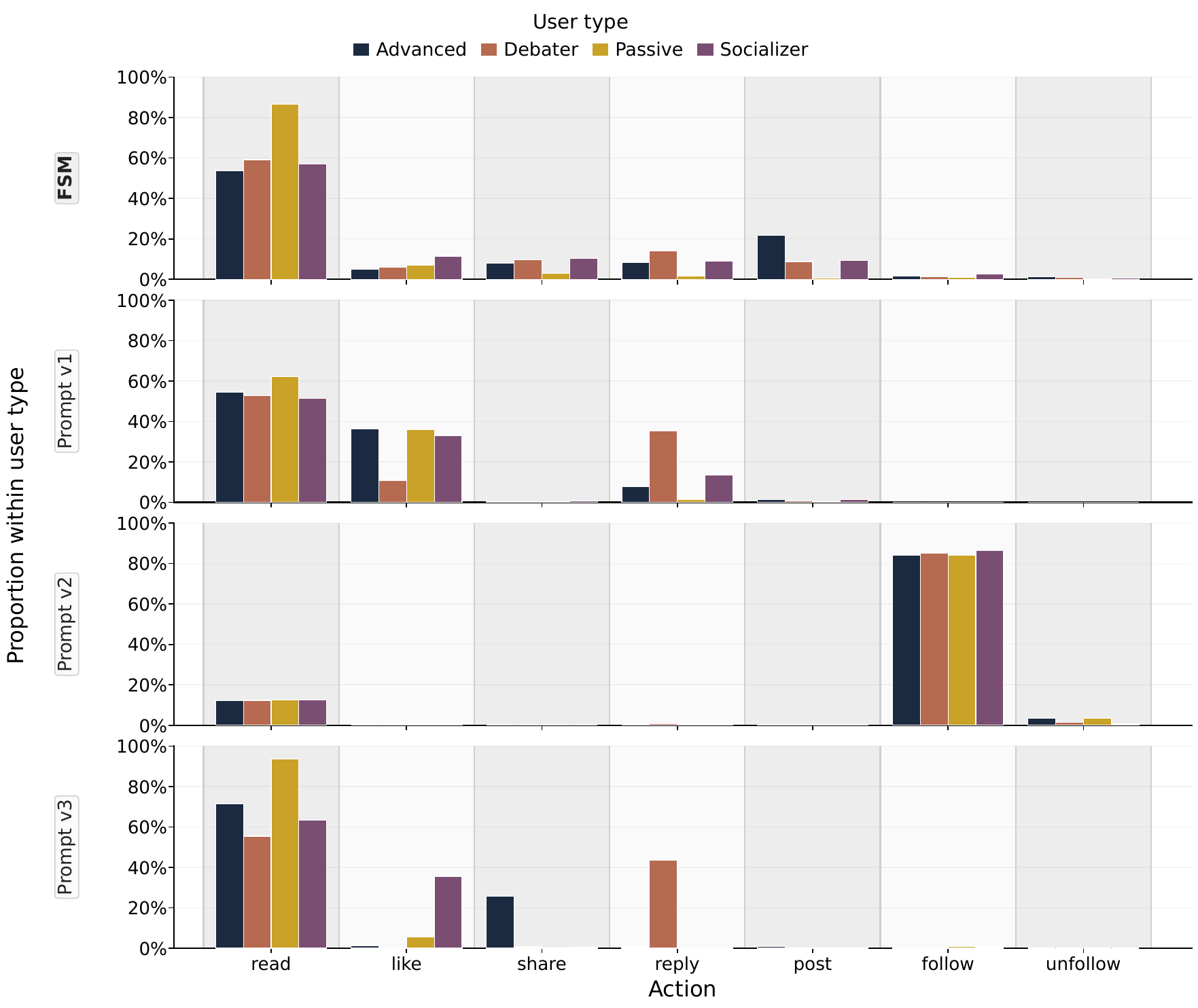}
\caption{Action distributions produced by GPT-OSS under the three prompting strategies, compared with the FSM baseline.}
\label{fig:bargpt}
\end{figure}

Figure~\ref{fig:bargpt} shows the action proportions produced by GPT-OSS under the three prompts, compared with the FSM baseline. Base prompt (v1) shifts part of the distribution toward \textit{like}, reducing several lower-frequency actions. Guided prompt (v2) departs most strongly from the FSM, with \textit{follow} dominating across user types. Probabilistic prompt (v3) restores \textit{read} for several user types and reduces the concentration on \textit{follow}, although some actions remain less represented.

\begin{figure}[ht!]
\centering
\includegraphics[width=\linewidth]{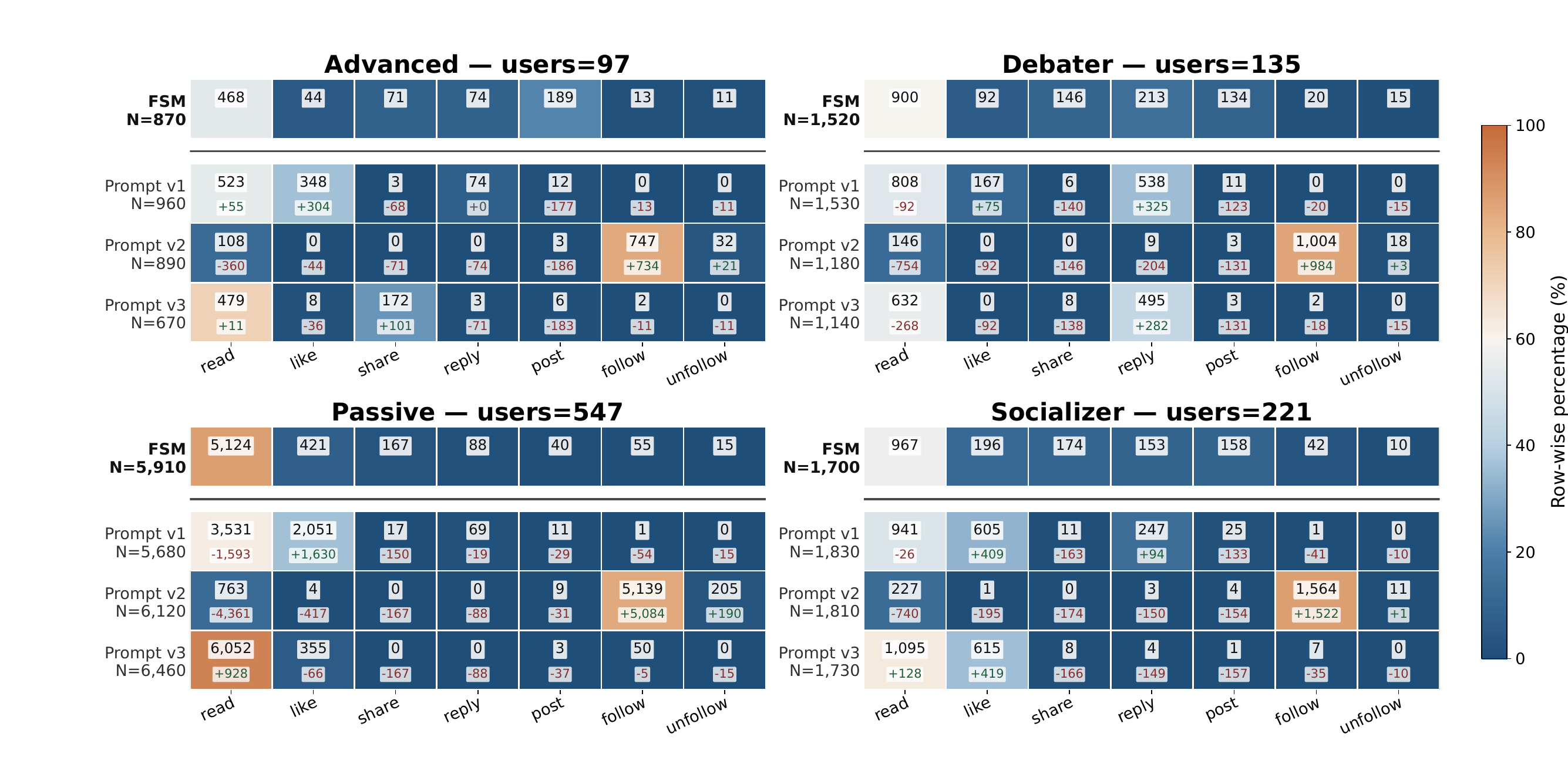}
\caption{Action counts produced by GPT-OSS by user type and prompt.}
\label{fig:heatgpt}
\end{figure}

Figure~\ref{fig:heatgpt} complements the proportional view by showing the raw action counts and their differences with respect to the FSM. Base prompt (v1) shifts many decisions toward \textit{like} while reducing several less frequent actions, remaining relatively closer to a consumption-oriented distribution. Guided prompt (v2) produces a strong concentration on \textit{follow}, with this action dominating the counts across user types. Probabilistic prompt (v3) reduces this concentration and restores \textit{read} as the most frequent action, although differences with respect to the FSM remain for some actions.

The JSD results are consistent with the distributional differences observed in the figures. At the global level, probabilistic prompt (v3) achieves the closest alignment with the FSM ($JSD=0.035$), followed by base prompt (v1) ($JSD=0.113$), whereas guided prompt (v2) shows the largest divergence ($JSD=0.672$). The weighted user-type averages follow the same pattern, with v3 obtaining the lowest divergence ($JSD=0.087$), v1 remaining moderately aligned ($JSD=0.120$), and v2 remaining highly divergent ($JSD=0.676$). However, this alignment is not uniform across user types. Probabilistic prompt (v3) is closest for Passive users ($JSD=0.027$), but shows higher divergence for Advanced users ($JSD=0.180$), Debaters ($JSD=0.166$), and Socializers ($JSD=0.179$). In contrast, guided prompt (v2) exhibits high divergence across all user types, with values ranging from $0.660$ to $0.686$.

\subsection{Mistral 24B}
\label{sec-mistral}

\begin{figure}[ht!]
\centering
\includegraphics[width=\linewidth]{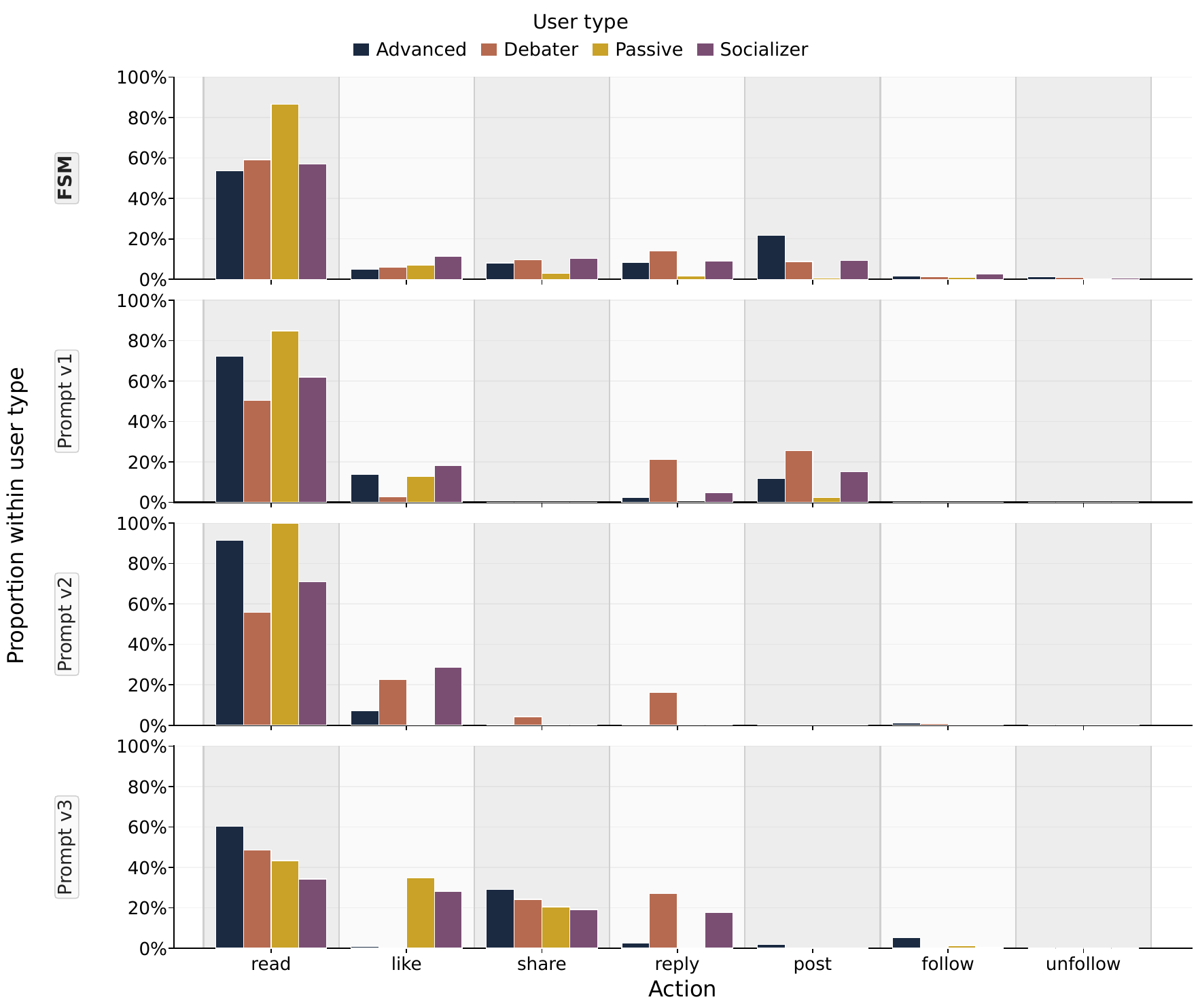}
\caption{Action distributions produced by Mistral 24B under the three prompting strategies, compared with the FSM baseline.}
\label{fig:barmistral}
\end{figure}

Figure~\ref{fig:barmistral} shows the action proportions produced by Mistral 24B under the three prompts, compared with the FSM baseline. Base prompt (v1) preserves \textit{read} as the dominant action across user types and does not concentrate the distribution on a single alternative action. However, some differences with respect to the FSM appear, including reduced representation of lower-frequency actions such as \textit{share}, \textit{follow}, and \textit{unfollow}. Guided prompt (v2) further increases the proportion of \textit{read}, producing a narrower distribution across actions. This effect is particularly visible for Passive users, where \textit{read} accounts for most decisions. Probabilistic prompt (v3) produces a broader distribution by increasing \textit{like} and \textit{share}, while reducing the dominance of \textit{read} for some user types, especially Passive users, resulting in a distribution that differs more from the FSM baseline.

\begin{figure}[ht!]
\centering
\includegraphics[width=\linewidth]{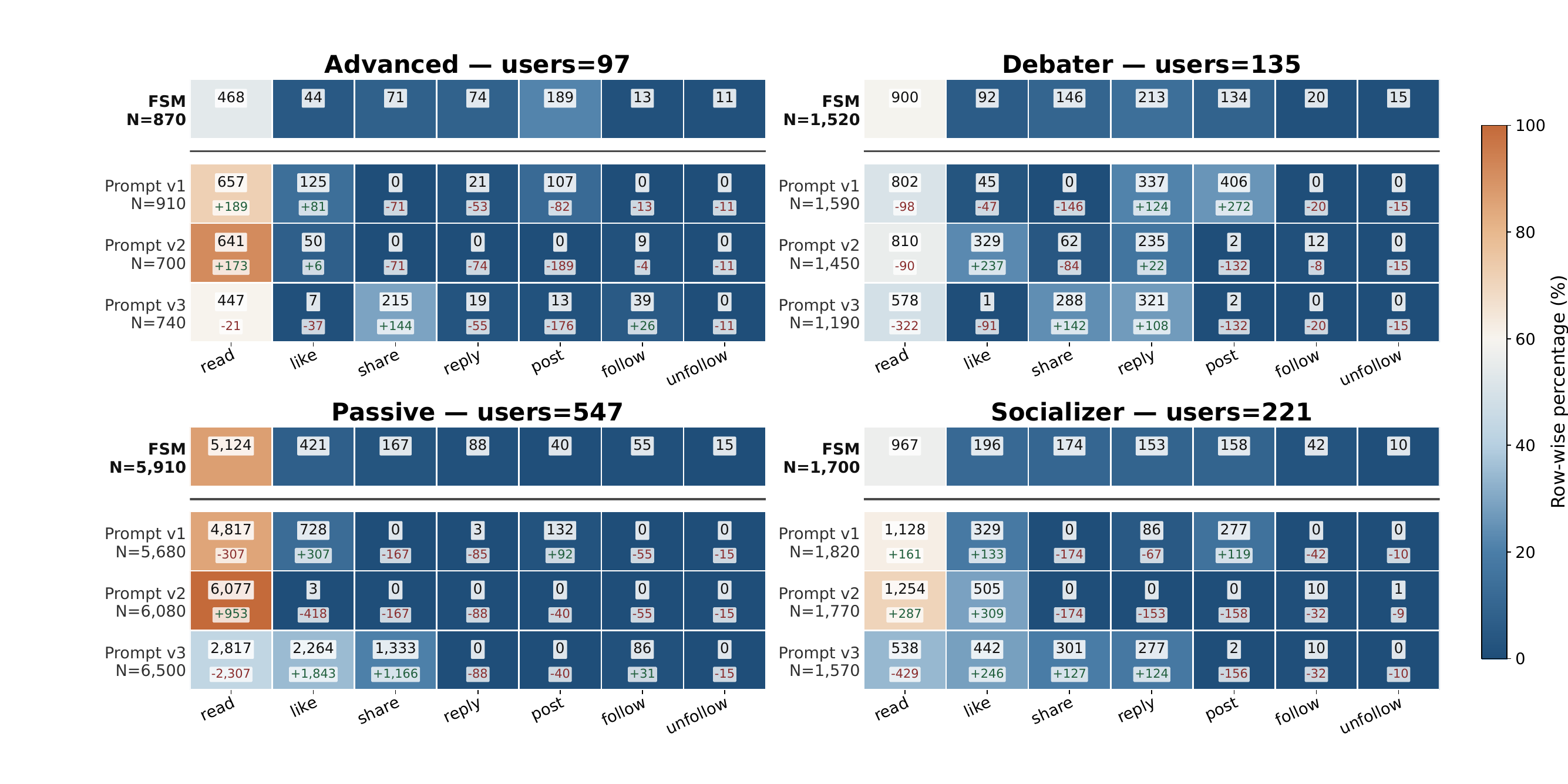}
\caption{Action counts produced by Mistral 24B by user type and prompt.}
\label{fig:heatmistral}
\end{figure}

Figure~\ref{fig:heatmistral} shows the count-level differences underlying these patterns. Base prompt (v1) remains relatively close to the FSM, mainly because it preserves the dominance of \textit{read}, although it reduces several less frequent actions. Guided prompt (v2) further concentrates decisions on \textit{read}, particularly for Passive users. Probabilistic prompt (v3) redistributes decisions across a wider set of actions by increasing \textit{like} and \textit{share}, while reducing \textit{read} for some user types.

The JSD results are consistent with these distributional patterns. At the global level, base prompt (v1) achieves the closest alignment with the FSM ($JSD=0.045$), followed closely by guided prompt (v2) ($JSD=0.055$), whereas probabilistic prompt (v3) shows higher divergence ($JSD=0.132$). The weighted user-type averages follow the same ordering, with v1 obtaining the lowest divergence ($JSD=0.060$), followed by v2 ($JSD=0.105$) and v3 ($JSD=0.165$). At the profile level, v1 shows lower divergence for Advanced users, Passive users, and Socializers, while v2 is slightly closer for Debaters. In contrast, v3 shows higher divergence for Passive users, indicating that changes in the balance between actions affect alignment for this profile.

\subsection{Cross-Method Comparison}
\label{sec:results_cross_model}

Table~\ref{tab:cross_model_results} summarizes global action-distribution alignment and execution cost of all configurations. Here, global alignment is computed as JSD over the full action distribution after the $10{,}000$ action-selection steps, without separating user types. Lower JSD values indicate that the LLM-selected actions are closer to the FSM reference distribution. Execution time reports the cost of selecting the same $10{,}000$ actions. The FSM is included as the reference policy: it has zero divergence by definition and only requires direct sampling from the transition matrix.

\begin{table}[ht!]
\centering
\caption{Global distributional alignment and execution cost across decision policies and models. Execution time is computed over $10{,}000$ action-selection steps. Bold values indicate the best LLM alignment and the fastest LLM execution time.}
\label{tab:cross_model_results}
\scriptsize
\setlength{\tabcolsep}{4.5pt}
\renewcommand{\arraystretch}{1.12}

\sisetup{
    detect-weight=true,
    detect-inline-weight=math,
    output-decimal-marker={.},
    group-separator={,},
    group-minimum-digits=4,
    group-digits=integer
}

\begin{tabular}{
    l
    l
    !{\vrule width 0.35pt}
    S[table-format=1.3]
    S[table-format=4.0]
    S[table-format=1.4]
    S[table-format=4.1]
}
\toprule
\textbf{Decision policy} &
\textbf{Model} &
\multicolumn{1}{c}{\textbf{JSD}} &
\multicolumn{3}{c}{\textbf{Execution cost}} \\
\cmidrule(lr){4-6}
& & &
{\textbf{Time (s)}} &
{\textbf{s/action}} &
{\textbf{vs. FSM}} \\
\midrule

FSM & -- & 0.000 & 7 & 0.0007 & 1.0 \\

\midrule
\multirow{3}{*}{Prompt v1}
& LLaMA 3.1   & 0.359 & \bfseries 946  & \bfseries 0.0946 & \bfseries 135.1 \\
& GPT-OSS     & 0.113 & 2172 & 0.2172 & 310.3 \\
& Mistral 24B & 0.045 & 2989 & 0.2989 & 427.0 \\

\midrule
\multirow{3}{*}{Prompt v2}
& LLaMA 3.1   & 0.223 & 3089 & 0.3089 & 441.3 \\
& GPT-OSS     & 0.672 & 6334 & 0.6334 & 904.9 \\
& Mistral 24B & 0.055 & 9360 & 0.9360 & 1337.1 \\

\midrule
\multirow{3}{*}{Prompt v3}
& LLaMA 3.1   & 0.278 & 1816 & 0.1816 & 259.4 \\
& GPT-OSS     & \bfseries 0.035 & 2990 & 0.2990 & 427.1 \\
& Mistral 24B & 0.132 & 5794 & 0.5794 & 827.7 \\

\bottomrule
\end{tabular}
\end{table}

At the prompting-strategy level, the results show that LLM-based policies can approximate the FSM distribution in some configurations, but no prompting strategy is consistently best across models. Averaging JSD across the three models, probabilistic prompt (v3) obtains the lowest mean divergence ($0.148$), followed by base prompt (v1) ($0.172$), while guided prompt (v2) has the highest mean divergence across models ($0.317$). However, this average ranking masks model-specific effects: probabilistic prompt (v3) is best for GPT-OSS, base prompt (v1) is best for Mistral 24B, and guided prompt (v2) is best for LLaMA 3.1. These results show that adding behavioral or numerical guidance does not consistently lead to closer alignment with the FSM.

At the model level, Mistral 24B has the lowest average divergence from the FSM across prompts. Its mean JSD across prompts is $0.077$, clearly lower than GPT-OSS ($0.273$) and LLaMA 3.1 ($0.287$). GPT-OSS achieves the best single result with probabilistic prompt (v3) ($JSD=0.035$), but it also produces the worst configuration with guided prompt (v2) ($JSD=0.672$), indicating prompt sensitivity. LLaMA 3.1 is the fastest model, but its alignment remains weaker than Mistral 24B in all prompting strategies. The latency ranking differs from the alignment ranking: LLaMA 3.1 is the fastest model on average, followed by GPT-OSS and Mistral 24B. Therefore, fidelity and efficiency do not point to the same configuration: the model with the lowest average divergence is also the most expensive, while the fastest model does not achieve the lowest divergence from the FSM.

\section{Conclusion}
\label{sec:conclusion}

This paper examined whether LLM-based action selection preserves an interpretable finite-state decision policy in an agent-based OSN simulation. The comparison was conducted under identical agents, network, action space, and decision contexts, using three open-weight LLMs and three prompting strategies. Alignment with the FSM reference policy was measured using JSD, and execution time was reported.

The results provide a conditional answer to the research question. LLM-based selectors can approximate the FSM policy in some cases, but they do not preserve it consistently. Across the nine LLM configurations, the mean global JSD was $0.212$, which indicates that the average action distribution remains different from the FSM. However, some configurations achieved close alignment, showing that LLM-based action selection is not inherently unable to reproduce an explicit policy.

Prompt design alone was not sufficient to preserve the FSM policy, and it did not define a stable decision policy across models. Additional behavioral guidance or numerical transition information did not consistently improve alignment, and the same prompting strategy could move one model closer to the FSM while moving another further away. Therefore, LLM-based action selection should not be assumed to follow an intended policy solely because the prompt describes it.

The computational cost results were unambiguous. Across all LLM configurations, action selection was on average $563.3\times$ slower than the FSM, and even the fastest LLM configuration was $135.1\times$ slower. Replacing an explicit FSM/Markov policy with an LLM selector should therefore be treated as a modeling trade-off rather than a direct substitution. LLMs may add contextual flexibility, but replacing an explicit policy with an LLM changes both the computational cost and the effective decision mechanism.

Future work should connect this decision-level benchmark with full OSN simulations and study how action-level divergence affects diffusion, engagement, network evolution, and narrative amplification.






\begin{credits}
\subsubsection{\ackname} This work has been partially funded by the University of Murcia with the FPI/0000902983 contract.

\subsubsection{\discintname}
The authors have no competing interests to declare that are relevant to the content of this article.
\end{credits}

%
%
%
\bibliographystyle{splncs04}
\bibliography{biblio}
\end{document}